\documentclass[10pt,twocolumn,letterpaper]{article}

\usepackage[pagenumbers]{cvpr} 

\usepackage[dvipsnames]{xcolor}

\definecolor{cvprblue}{rgb}{0.21,0.49,0.74}
\usepackage[pagebackref,breaklinks,colorlinks,citecolor=cvprblue]{hyperref}
\usepackage[export]{adjustbox}
\usepackage{pgfplots}
\pgfplotsset{compat=newest}
\usetikzlibrary{3d,arrows.meta,shapes.geometric,calc}
\usepackage{tikz-3dplot}

\def\paperID{16822} 
\def\confName{CVPR}
\def\confYear{2024}

\title{Volumetric Reconstruction Resolves Off-Resonance Artifacts in Static and Dynamic 
PROPELLER MRI}

\author{Annesha Ghosh,$^1$ Gordon Wetzstein,$^2$ Mert Pilanci,$^2$ and Sara Fridovich-Keil$^2$\\
$^1$University of California, Berkeley \hspace{2cm} $^2$Stanford University\\
{\tt\small annesha.g@berkeley.edu, \{gordon.wetzstein, pilanci, sarafk\}@stanford.edu}
}

\begin{document}
\maketitle


\begin{abstract}

Off-resonance artifacts in magnetic resonance imaging (MRI) are visual distortions that occur when the actual resonant frequencies of spins within the imaging volume differ from the expected frequencies used to encode spatial information. 
These discrepancies can be caused by a variety of factors, including magnetic field inhomogeneities, chemical shifts, or susceptibility differences within the tissues. 
Such artifacts can manifest as blurring, ghosting, or misregistration of the reconstructed image, and they often compromise its diagnostic quality.
We propose to resolve these artifacts by lifting the 2D MRI reconstruction problem to 3D, introducing an additional ``spectral'' dimension to model this off-resonance. 
Our approach is inspired by recent progress in modeling radiance fields, and is capable of reconstructing both static and dynamic MR images as well as separating fat and water, which is of independent clinical interest.
We demonstrate our approach in the context of PROPELLER (Periodically Rotated Overlapping ParallEL Lines with Enhanced Reconstruction) MRI acquisitions, which are popular for their robustness to motion artifacts.
Our method operates in a few minutes on a single GPU, and to our knowledge is the first to correct for chemical shift in gradient echo PROPELLER MRI reconstruction without additional measurements or pretraining data.
\end{abstract}

\section{Introduction} \label{sec:intro}
\newcommand{\plotright}[1]{%
  \adjincludegraphics[trim={{0.5\width} {0\height} {0\width} {0\height}}, clip, width=0.24\linewidth]{#1}%
}
\newcommand{\plotleft}[1]{%
  \adjincludegraphics[trim={{0\width} {0\height} {0.5\width} {0\height}}, clip, width=0.24\linewidth]{#1}%
}
\newcommand{\plottrim}[1]{%
  \adjincludegraphics[trim={{0.05\width} {0.05\height} {0.05\width} {0.05\height}}, clip, width=0.24\linewidth]{#1}%
}
\begin{figure}[h]
\includegraphics[width=\linewidth]{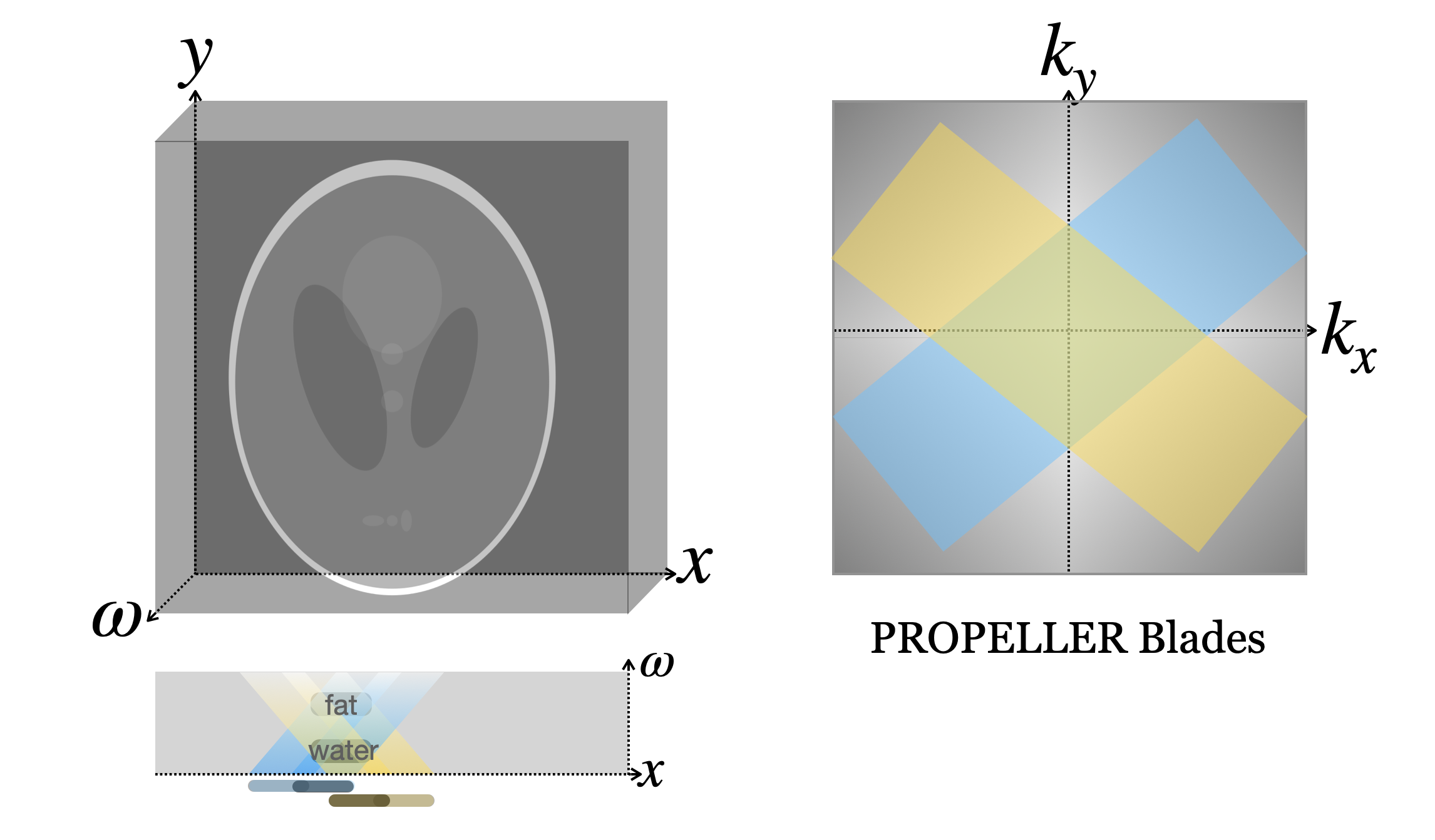}
\plotleft{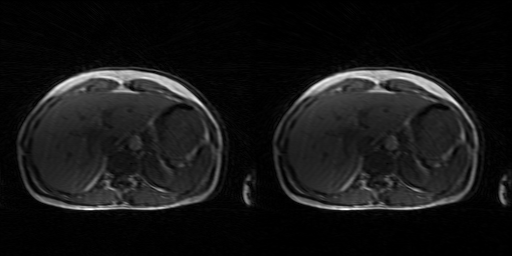}
\plotleft{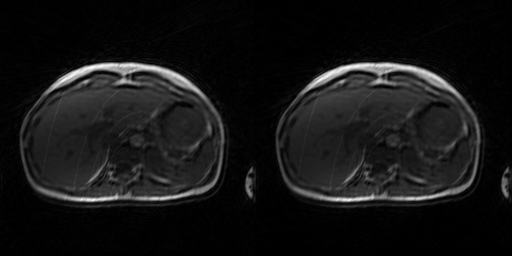}
\plottrim{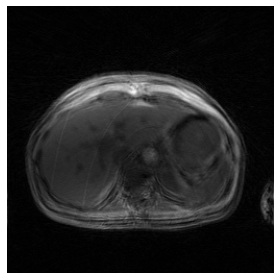}
\plotright{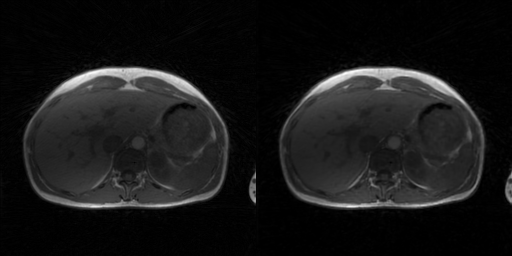}
\centering\small{~~~~~~~~~~~~~~2 of 5 Total Training Views~~~~~~~~~~~~Baseline~~~~~~~~~~~~~~~~Our~~~~~~}

\centering\small{~~~~~~~~~~~~(DFT$^{-1}$ of Blades)~~~~~~~~~~~~Reconstruction~~~Reconstruction}
\caption{\textbf{Overview of our method, shown here for reconstructing a static image.} On the top left, we model a 2D magnetic resonance image (here, the Shepp--Logan phantom) as a 3D volume in real-space, using a third dimension for frequency shift to account for off-resonance. On the right, we illustrate two PROPELLER blades in $k$-space, with measurement support highlighted in blue and yellow. The cross-section shows the off-resonance (chemical shift) effects produced by these two blades, which result in offset images of fat and water. By optimizing a 3D model rather than a 2D image, we can produce an uncorrupted 2D reconstruction (bottom right) by performing a projection exactly aligned with the spectral axis, shown here for an abdomen cross-section focusing on the liver.} \label{fig:method}
\end{figure}

Magnetic resonance imaging (MRI) is an essential tool for biomedical research as well as noninvasive diagnosis and monitoring of a wide variety of medical conditions. MRI does not require any ionizing radiation, making it safer than imaging with X-rays, and can be used to measure both bone and soft tissues. However, MRI measurements are slow to collect, which limits their ability to image dynamic bodily functions like peristalsis, respiration, and the cardiac cycle.
We therefore consider PROPELLER (Periodically Rotated Overlapping ParallEL Lines with Enhanced Reconstruction) MRI, which is recognized for its ability to correct for motion artifacts and enable dynamic MR imaging. 
PROPELLER MRI works by collecting $k$-space measurements in ``blades'' that overlap in the low frequencies, as illustrated in the top right of \Cref{fig:method} and in the top row of \Cref{fig:input}. Each blade can be sampled relatively quickly, and by combining different blades sampled over time it is possible to recover a dynamic MR image.

However, gradient-echo PROPELLER reconstructions suffer from off-resonance or chemical shift artifacts, in which the distribution of fat and water in a sample causes local distortions of the magnetic field. These distortions appear in the reconstructed images as a translation of fat and water relative to each other, with different blade measurements in $k$-space producing different translation artifacts in image space. Left uncorrected, off-resonance artifacts limit our ability to combine the data from different blades, and can obscure the true spatial location of affected tissues, as shown in the baseline reconstruction in \Cref{fig:method}. Currently, the only way to perform PROPELLER MRI without chemical shift artifacts is to use slower spin echo measurements, or estimate a magnetic field map using additional samples, either of which limit the temporal resolution of the imaging. Our goal is to enable chemical shift correction for fast gradient-echo PROPELLER MRI without additional measurements, which may open the door to faster scan times and higher temporal resolution of dynamic functions like heartbeat, respiration, and peristalsis.

We combine the partial reconstructions from each gradient-echo PROPELLER blade to estimate the off-resonance and produce a single fused image or video that is free of off-resonance artifacts. We achieve this by introducing an additional ``spectral'' dimension that models how much the tissue at each spatial location is susceptible to chemical shift. The contents of each pixel are allowed to exist at different ``heights'' in this extra dimension, where the height of a structure represents its off-resonance and the amount by which it shifts in 2D between the reconstructions of different PROPELLER blades. 

Specifically, we show that the 2D reconstruction from a PROPELLER blade at angle $\theta$ corresponds to a projection of our 3D model at the same angle. If the contents of a 2D location, or pixel, exist at different heights in our model -- corresponding to different off-resonance effects -- they will project to different 2D locations relative to each other. As the blade angle $\theta$ changes, so will the 2D offset between these components in the corresponding projection of our model. Since the magnetic field inhomogeneity is a function of the sample being measured, and not of the $k$-space sampling trajectory, a single 3D volume must be consistent with all of our PROPELLER blade reconstructions. We therefore \emph{optimize} the contents of our 3D grid to be consistent with our PROPELLER measurements, and render out a fused image without off-resonance artifacts by projecting our 3D model along its spectral dimension. 

Our approach demonstrates successful correction of chemical shift effects in both static and dynamic PROPELLER MRI, using a mixture of real datasets and synthetic datasets that mimic the effects in real tissue and real gradient echo measurements. 
To our knowledge, ours is the first method to correct for off-resonance artifacts in gradient-echo PROPELLER MRI without additional measurements, and we are hopeful that our work opens the door to faster and higher-fidelity imaging of organ motion.


\section{Related Work} \label{sec:related}


\subsection{PROPELLER MRI}

PROPELLER (Periodically Rotated Overlapping ParallEL Lines with Enhanced Reconstruction) magnetic resonance imaging (MRI) \cite{forbes2001propeller} is the $k$-space sampling trajectory for which we aim to correct off-resonance artifacts, separate fat and water, and resolve motion. The PROPELLER trajectory is illustrated in the top right of \Cref{fig:method} and the top row of \Cref{fig:input}, and consists of overlapping ``blades'' through the origin in $k$-space. One blade is collected per timestep, so that all of $k$-space is sampled at some time but the measurement at each timestep can be collected relatively quickly by undersampling high frequencies in that direction. Spin-echo PROPELLER MRI is popular for its robustness to motion artifacts, but faster gradient-echo PROPELLER MRI is currently unused because of its susceptibility to chemical shift; it is this susceptibility that we aim to resolve.


\subsection{Off-resonance correction}

Several approaches have been established to address off-resonance artifacts in various MRI acquisition patterns \cite{smith2010mri}; we summarize some of the most relevant methods and how they relate to our approach.


\paragraph{Field map estimation.}

A popular physics-based approach computes a field map to correct for the effects of magnetic field inhomogeneity in non-Cartesian sampled MRI by estimating a spatially varying sample density compensation function \cite{noll2005conjugate}. 
Field map estimation can be done more efficiently by leveraging compressed sensing \cite{doneva2010compressed}, but still requires some additional $k$-space measurements.

\paragraph{IDEAL method for spin-echo MRI.}

The Iterative Decomposition of water and fat with Echo Asymmetry and Least-squares estimation (IDEAL) method \cite{reeder2005iterative} addresses off-resonance for spin-echo MRI by combining asymmetrically acquired echoes using an iterative algorithm.
This approach is effective for spin-echo measurements, but these are slower to acquire than gradient-echo measurements which are the focus of our work.



\paragraph{Frequency segmented correction.}
Frequency segmented off-resonance correction works by modulating the raw measurements by different frequencies to produce base images in which local off-resonance effects appear as local blurring. This local blurring is then estimated using a focus metric, and the in-focus regions from each base image are combined into a single uncorrupted image. 
Frequency segmented correction has been demonstrated for spiral $k$-space sampling trajectories \cite{noll1992deblurring} as well as with an available field map estimate \cite{man1997multifrequency}.
We focus instead on the PROPELLER $k$-space sampling trajectory, for which off-resonance artifacts appear as local spatial shifts rather than local blurring, and we do not assume access to a field map estimate.





\paragraph{Deep Learning.}

Deep learning approaches have also been explored as a means to correct for chemical shifts. For example, deep learning methods using a residual convolutional neural network have been trained to predict uncorrupted magnetic resonance images in 3D conical MRI \cite{zeng2019deep} as well as spiral RT-MRI \cite{lim2020deblurring}. 
These methods can work without access to field maps, but instead require access to pretraining examples of corrupted and clean magnetic resonance images. 
This pretraining data may be cumbersome to acquire, and deep models trained on them may only perform well on similar data.
Our method does not require any pretraining data, and instead leverages the physical equivalence between chemical shift and volumetric projection, as described in \Cref{sec:formulation}.



\paragraph{Dixon-RAVE.}
So-called ``Dixon'' methods correct for off-resonance using shift encoding, allowing them to separate fat and water using Cartesian-sampled measurements taken with different echo times. Most similar to ours is Dixon-RAVE \cite{dixon-rave}, which extends Dixon correction to gradient echo non-Cartesian $k$-space sampling.
Both our method and Dixon-RAVE share the ability to correct gradient-echo MRI, recover motion, and operate in the absence of field map estimates or pretraining data. However, the two methods differ in that Dixon-RAVE focuses on radial sampling whereas we focus on PROPELLER sampling, and in that our method recovers continuous-valued spectral shifts rather than a binary separation of fat and water.
Additionally, Dixon-RAVE resolves dynamics by binning measurements into discrete states based on principal component analysis, whereas we reconstruct a video with each PROPELLER measurement assigned to its own frame. We expect that the binning approach may provide greater noise robustness for truly periodic motion, such as regular breathing, but our approach may be more flexible to one-off or irregular motions, such as peristalsis.


\subsection{Volumetric reconstruction}

Our inspiration for the use of 3D reconstruction comes from recent progress in modeling radiance fields, in particular the grid-based method Plenoxels \cite{fridovich2022plenoxels}. Plenoxels optimizes a 3D volume from 2D photographs using backpropagation through batches of camera rays, with samples along each ray evaluated using trilinear interpolation of the neighboring voxels. We use a similar strategy for evaluating projections and optimizing a volume, which in our case represents a two-dimensional static image with a third spectral dimension to model off-resonance. 

For dynamic MRI off-resonance correction we likewise leverage recent progress in compressive modeling of dynamic scenes, in particular K-Planes \cite{kplanes_2023}. 
This representation stores parameters in planes over pairs of dimensions, and is a dynamic extension of the triplane volume representation introduced in EG3D \cite{eg3d}. 

We adapt both static and dynamic scene representations from radiance fields to MRI by removing color and view-dependent effects, and instead representing complex values at each spatial or spatio-temporal location. Additionally, instead of the volume rendering formula used for camera measurements, the corresponding formula for MRI is a direct, linear projection as illustrated in \Cref{fig:shift_illustration}.


\section{Mathematical Formulation} \label{sec:formulation}



\begin{figure}
\input{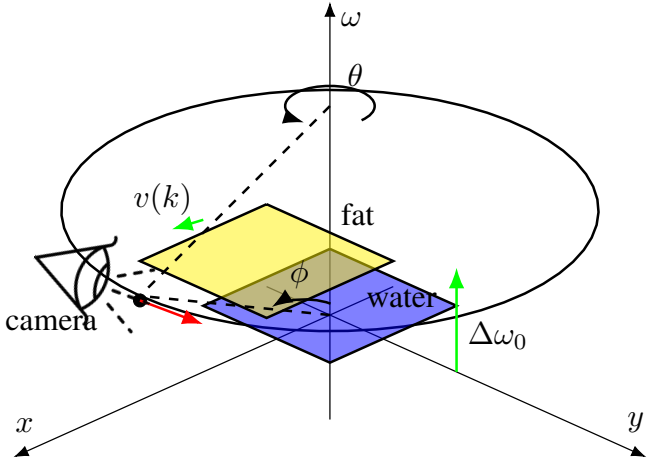}
\caption{\textbf{Diagram illustrating the equivalence of chemical shift and 3D projection.} Our model for a 2D MR image is a 3D volume with the two spatial dimensions $x$ and $y$ along with an extra spectral dimension $\omega$ that represents off-resonance.\vspace{-0.3cm}}
\label{fig:shift_illustration}
\end{figure}

Magnetic resonance $k$-space measurements $s$ in the presence of chemical shift are defined by the signal equation \cite{nishimura_book, larson_ebook}:
\begin{equation}
    s(t) = \int m(\textbf{r})e^{-i\Delta \omega(\textbf{r})t}e^{-i\mathbf{k}(t)\cdot \mathbf{r}}d\mathbf{r}
\end{equation}
where $\Delta \omega$ denotes magnetic field inhomogeneity in radians, the source of off-resonance and chemical shift artifacts. Here $m(\textbf{r})$ is the real-space uncorrupted image and $\textbf{k}(t)$ is its $k$-space sampling trajectory over time $t$.
In the case of a single frequency shift $\Delta \omega(\textbf{r}) = \Delta \omega_0$, we can gain some intuition by simplifying the signal equation as follows:
\begin{align*}
    s(t) &= e^{-i\Delta\omega_0 t}\int m(\textbf{r})e^{-i\mathbf{k}(t)\cdot \mathbf{r}}d\mathbf{r} \\
    &= M(\textbf{k}(t))e^{-i\Delta\omega_0 t}.
\end{align*}
In this form, the frequency shift manifests as a phase accumulation as the $k$-space measurements are collected over time. We can write this succinctly using the notation $v(\textbf{k})$ to track the $k$-space sampling trajectory which controls this phase accumulation:
\begin{align*}
    \hat M(\textbf{k}(t)) &:= M(\textbf{k}(t))e^{-i\Delta\omega_0 v(\textbf{k})} .
\end{align*}
If we treat this corrupted $k$-space measurement as our Fourier-domain signal $\hat M$, its inverse Fourier transform produces a corrupted image $\hat m$ that exhibits chemical shift. For a single frequency shift $\Delta \omega_0$ this takes the form of a convolution of the uncorrupted image $m$ with the accumulated phase:
\begin{equation}
\label{eqn:convolution}
    \hat m(\textbf{r}) = m(\textbf{r}) * \mathcal{F}^{-1}(e^{-i v(\textbf{k})\Delta \omega_0}).
\end{equation}
In the case of PROPELLER acquisitions, $v(\textbf{k})$ is a unit vector aligned with the PROPELLER blade. 
Therefore, due to the shift property of the Fourier transform, a single frequency shift $\Delta \omega_0$ results in a global translation of the true image $m$ in the direction of the PROPELLER blade.
This effect is illustrated in \Cref{fig:shift_illustration}, where the global translation is modeled as a perspective shift in an off-axis projection from 3D to 2D.

In reality, each pixel experiences its own frequency shift $\Delta \omega(\textbf{r})$ determined by the material properties, namely fat and water content, of the tissue at that pixel location. \Cref{eqn:convolution} then applies to each material in the uncorrupted image $m(\textbf{r})$, which causes regions of fat and water to shift locally relative to each other. 
As shown in \Cref{fig:shift_illustration}, \Cref{eqn:convolution} is equivalent to an off-axis projection of a volume along the spectral dimension, where the ``height'' of fat relative to water is $\Delta \omega_0$.
We resolve this local chemical shift by modeling a continuous spectral dimension that allows each spatial location to experience a continuous-valued chemical shift effect. This models the reality that different tissues contain a continuous spectrum of chemical properties and corresponding chemical shift values $\Delta \omega$.

\section{Method} \label{sec:method}

The input to our method is a set of $k$-space PROPELLER blade measurements and their corresponding real-space 2D images, obtained via the inverse DFT. Examples of such images are shown in \Cref{fig:input} and \Cref{fig:real_data_input}. 

Each training image is a reconstruction of the same underlying structure, but each image exhibits two visual artifacts. One artifact is a directional blur corresponding to the $k$-space PROPELLER blade, in which high frequencies in a specific orientation are not sampled. The other artifact is due to off-resonance, in which regions of fat (bright) and water (dim) appear offset from each other in the image, compared to their true positions in space. As the angle of the PROPELLER blade rotates, so does the direction of the blur as well as the direction of the fat-water shift. The blur artifacts are more easily recovered, as the blur is determined solely by the $k$-space sampling pattern and is therefore known \emph{a priori}. However, the off-resonance shift artifacts are data specific, determined by the arrangement of fat and water which is unknown \emph{a priori}.

We correct for both of these artifacts by lifting the 2D image reconstruction task into a 3D volume reconstruction task, in which we can account for data-dependent chemical shifts. Our method contains the following elements, illustrated in \Cref{fig:method}. Our code is available at \url{github.com/sarafridov/volumetric-propeller}.

\subsection{Volume model}
The optimizable parameters in our model represent the complex-valued 3D volume over $x, y$, and the spectral dimension $\omega$, shown in the top left of \Cref{fig:method}. For static imaging we arrange parameters directly in a 3D grid. For dynamic imaging we use a complex-valued K-Planes representation \cite{kplanes_2023} over the $x, y, \omega$, and $t$ (time) dimensions. The parameters of this dynamic model are stored in six complex-valued grids $\textbf{p}_{xy}$, $\textbf{p}_{x\omega}$, $\textbf{p}_{xt}$, $\textbf{p}_{y\omega}$, $\textbf{p}_{yt}$, and $\textbf{p}_{\omega t}$, as well as a complex-valued weighting vector $\textbf{a}$. 
The signal value at a spatio-temporal location $(x, y, \omega, t)$ is computed by projecting this 4D coordinate into each of the six grids and extracting a feature vector by bilinear interpolation in each 2D grid. These six feature vectors are then multiplied together elementwise (Hadamard product) and decoded into a single complex number via an inner product with the weighting parameter $\textbf{a}$. This process is summarized in \Cref{eqn:kplanes}:
\begin{equation}
\label{eqn:kplanes}
    f(x, y, \omega, t) = \langle \textbf{a}, \Pi_{c \in C}\psi(\textbf{p}_{c}, \pi_c(x, y, \omega, t)) \rangle,
\end{equation}
where $C = \{xy, x\omega, xt, y\omega, yt, \omega t\}$ denotes the set of six feature grids, $\pi_c$ denotes projection onto the 2D plane defined by dimensions $c$, and $\psi$ denotes bilinear interpolation of continuous-valued 2D coordinates within that plane. 

Each of these volume models---static and dynamic---performs interpolation between grid values so that the modeled function is continuous across space and time, with spatio-temporal bandwidth determined by the resolution in each dimension $x, y, t$. These resolution parameters are design variables that control the resolution of the final reconstruction. The resolution of the spectral dimension $\omega$ controls the severity of off-resonance artifacts that can be corrected for, and is generally much smaller than the spatial resolution in $x$ and $y$. 
For our dynamic model, there is an additional feature dimension for each 2D grid and the weighting vector $\textbf{a}$, which we set to 8.
This dynamic K-Planes model also supports a multiscale representation in which the parameters are replicated at different resolutions and the results are summed; we use two scales where one is the full image resolution and the other is half the resolution in the spatial dimensions.

%

\subsection{Forward model}

We render our volume model into a simulated magnetic resonance measurement using a two-step process: projection and blurring.

\paragraph{Projection.}
We use projection to generate a 2D image from our volume model, as illustrated in the cross-section in the middle row of \Cref{fig:method} as well as in \Cref{fig:shift_illustration}. We construct a ray corresponding to each image pixel, sample 3D points along each ray (in the three dimensions $x$, $y$, and $\omega$), evaluate each sample by interpolation into our volume model, and sum the samples along each ray to produce a final projected pixel value. 

A projection in 3D can be described by two angles, $\theta$ in the spatial image plane and $\phi$ with respect to the spectral axis. Each PROPELLER blade has an angle $\theta$ determined by the orientation of the blade in $k$-space, and we assign it a small positive angle $\phi$, such as $20^\circ$, so that the projections for different blades will differ. This procedure models off-resonance artifacts: two pieces of material at the same position in the image plane but with different spectral ``heights'' project to different image locations, with their relative shift varying according to the projection angle $\theta$ corresponding to each PROPELLER blade. This equivalence between off-resonance and projection is illustrated in \Cref{fig:shift_illustration}.

\paragraph{Blur.}
Each PROPELLER blade measurement is undersampled in $k$-space by omitting high frequencies along a certain direction. This $k$-space undersampling results in a directional blur in image space, shown in the training views at the bottom left of \Cref{fig:method} and in \Cref{fig:input}.

To simulate these measurements, we take each of our projected images, compute its 2D discrete Fourier transform (DFT), and then mask the DFT by the appropriate PROPELLER blade support in $k$-space. The resulting masked values can be used in $k$-space as simulated raw PROPELLER blade measurements, or moved back to image space by an inverse DFT.



\subsection{Optimization}
We optimize the grid values of our static and dynamic volume models by iterative gradient updates, using the Adam optimizer. At each iteration, we choose a PROPELLER blade, with a known $k$-space mask (at angle $\theta$) and corresponding measurement. We then apply our forward model and compute the square $L_2$ norm of the error between our prediction and the true measurement, in both $k$-space and image space. We sum these two data-fidelity terms, with most weight on the image-space loss, but include both terms to minimize any artifacts induced by the discretized $k$-space masking procedure. We also include total variation and $L_1$ sparsity regularization along the spatial and spectral dimensions of our volume model, which helps to produce clean and sparse edges and remove any initialization values in undersupervised corners of the volume model. For dynamic reconstruction, we also include temporal smoothness and sparsity regularizers to encourage motions to be smooth and limited in scope. After optimization is completed, we render a test image or video without off-resonance artifacts by projecting our volume model with $\phi=0$, so that material at the same position in the spatial plane projects to the same image location. For dynamic reconstruction, we do a final denoising step on each frame using a nonlocal means filter, to remove some grid artifacts induced by the tensor decomposition representation of the volume model.

\section{Datasets} \label{sec:datasets}


We validate our approach on four datasets, two static abdomen cross-sections (through liver and breast tissue, respectively) based on real magnetic resonance measurements provided publicly by Dixon-RAVE \cite{dixon-rave}, and two synthetic datasets based on the Shepp--Logan phantom, one static and another dynamic.
For each dataset, we begin with real-space complex-valued images or videos of fat and water layers, and combine these into a ground truth volume by embedding the fat and water layers along the spectral dimension of the volume. 
We then use our forward model to simulate PROPELLER measurements of this model that we use as input to our reconstruction method.

\paragraph{Static Shepp--Logan dataset.}

Our static Shepp--Logan dataset is defined by a complex-valued 3D slab with spatial dimensions $x$ and $y$ and an added spectral dimension $\omega$. 
We separate the standard Shepp--Logan phantom into two components, with the bright outer ellipse simulating fat and its interior simulating water-based tissue. We embed each of these layers in our 3D slab following a shallow Gaussian contour, so that there is small, smooth spectral variation within each layer and a larger spectral jump between the water layer and the fat layer. This setup is designed to recapitulate the reality that fat exhibits noticeably larger chemical shift artifacts compared to water, but there are also smaller spectral differences between different tissues of each type.

Using this synthetic ground truth volume, we produce training data by evaluating our forward model for each of five blades that together tile to fill a decagon centered at the origin in $k$-space. These $k$-space blades and the corresponding training images are shown in \Cref{fig:input}.

\newcommand{\plotleftsmall}[1]{%
  \adjincludegraphics[trim={{0\width} {0\height} {0.5\width} {0\height}}, clip, width=0.19\linewidth]{#1}%
}
\newcommand{\plotall}[1]{%
  \adjincludegraphics[trim={{0\width} {0\height} {0\width} {0\height}}, clip, width=0.19\linewidth]{#1}%
}
\begin{figure}[h]
\centering
\plotall{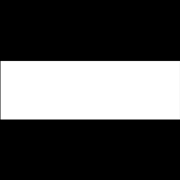}
\plotall{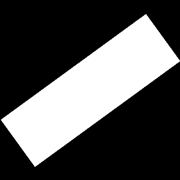}
\plotall{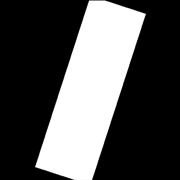}
\plotall{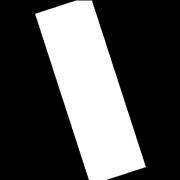}
\plotall{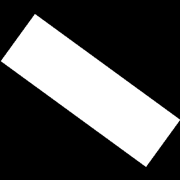}
\plotleftsmall{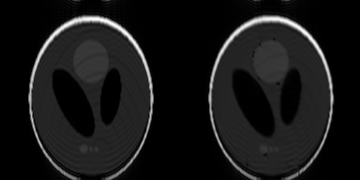}
\plotleftsmall{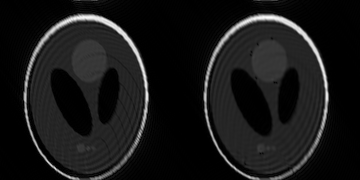}
\plotleftsmall{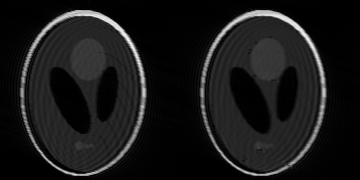}
\plotleftsmall{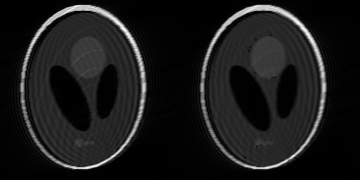}
\plotleftsmall{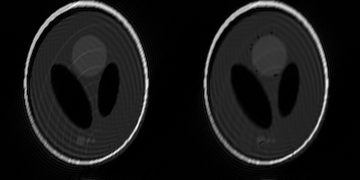}
\caption{\textbf{Static synthetic dataset: training images.} Top row: 5 $k$-space masks used to define the measurement blades. These 5 blades tile to cover a decagon centered at the origin in $k$-space. Bottom row: 5 input images corresponding to these blades. Note that the direction of chemical shift (the gap between the bright outer ellipse and the interior) and the direction of undersampling blur is determined by the orientation of the measurement blade in $k$-space. Each measurement amounts to viewing (projecting) the ground truth volume at an angle, and then blurring it.\vspace{-0.3cm}} \label{fig:input}
\end{figure}

\begin{figure}[h]
\centering
\plotleftsmall{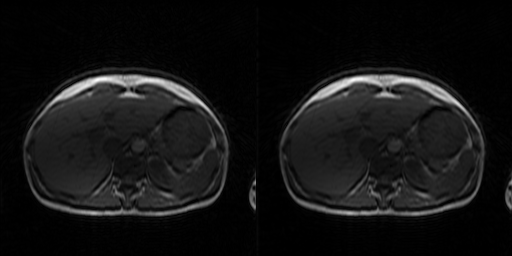}
\plotleftsmall{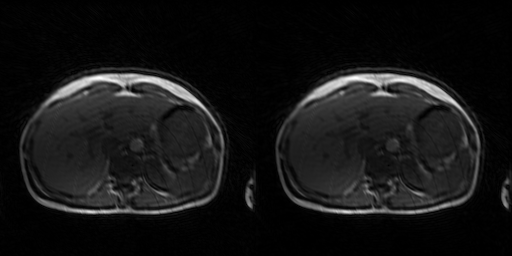}
\plotleftsmall{figures/liver/angle_3.png}
\plotleftsmall{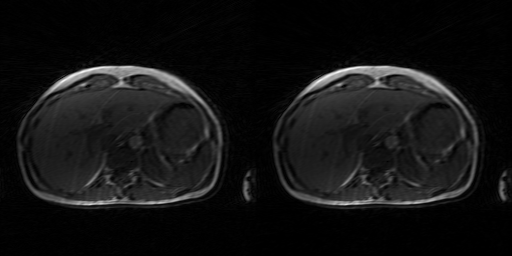}
\plotleftsmall{figures/liver/angle_1.png}
\plotleftsmall{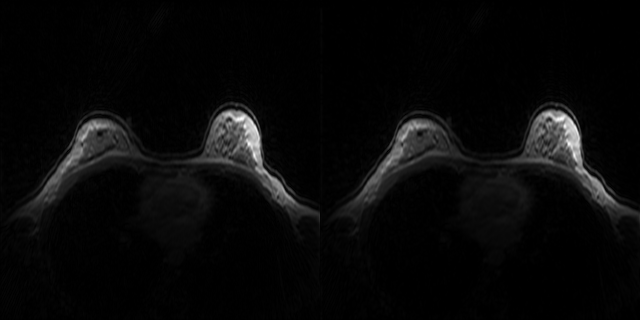}
\plotleftsmall{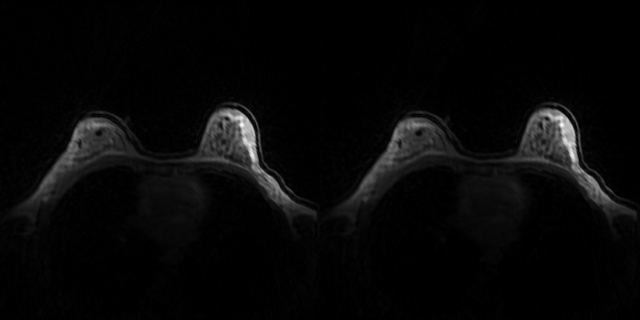}
\plotleftsmall{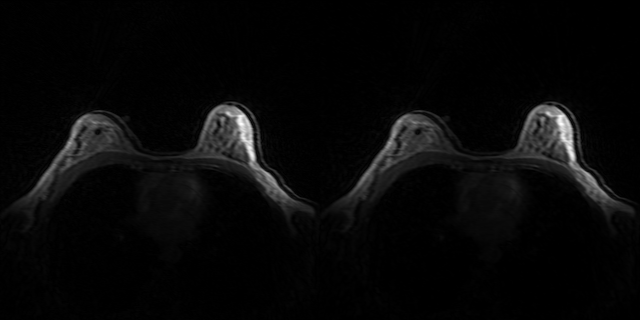}
\plotleftsmall{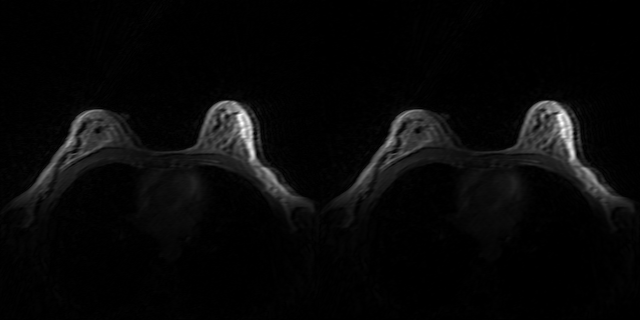}
\plotleftsmall{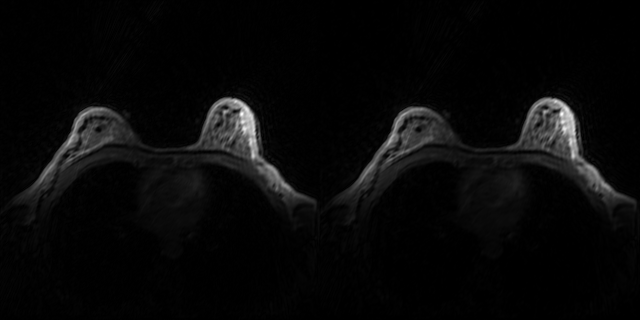}
\caption{\textbf{Static real datasets: training images.} Each row shows the 5 real input images corresponding to the 5 PROPELLER blades shown in the top row of \Cref{fig:input}. The top row shows our liver dataset and the bottom row shows our breast dataset, both of which are derived from the Dixon-RAVE \cite{dixon-rave} dataset.\vspace{-0.3cm}} 
\label{fig:real_data_input}
\end{figure}

\paragraph{Static liver and breast datasets.}
We use two static real-data magnetic resonance measurements, both based on public data provided by Dixon-RAVE \cite{dixon-rave}. This method separates free-breathing MRI reconstructions into complex-valued fat and water layers, which we then embed in our ground truth volume in the same manner as for the synthetic Shepp--Logan dataset. We then simulate PROPELLER $k$-space measurements using our forward model, which is necessary because the original data was not sampled according to the PROPELLER trajectory in $k$-space. Both of these real datasets visualize abdominal cross-sections, focusing on liver and breast tissue, respectively. Abdominal MRI is particularly susceptible to off-resonance artifacts because of the prevalence of both fat and water, and therefore is a particularly valuable application for our proposed off-resonance correction strategy. We show the training images for each of these abdominal datasets in \Cref{fig:real_data_input}.

\paragraph{Dynamic Shepp--Logan dataset.}

Our dynamic Shepp--Logan dataset is generated in much the same way as our static Shepp--Logan dataset, except that the original Shepp--Logan phantom is modified so that one of the ellipses moves along a sinusoidal pattern left and right over time. We collect a single $k$-space blade measurement at each time step, cycling among the five blades shown in \Cref{fig:input} for a total of 67 timesteps. 

\begin{figure}[h]
\centering
\plotleftsmall{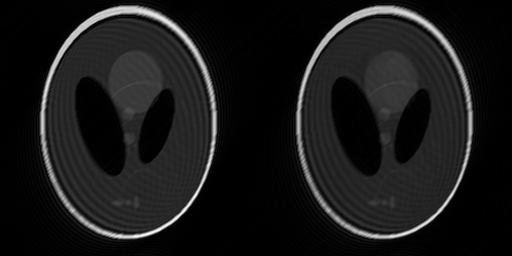}
\plotleftsmall{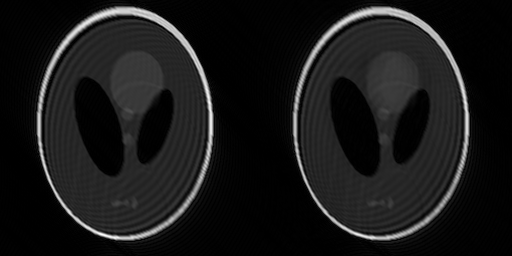}
\plotleftsmall{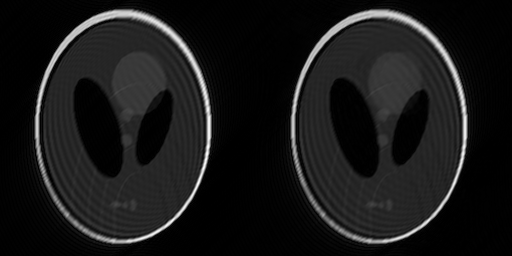}
\plotleftsmall{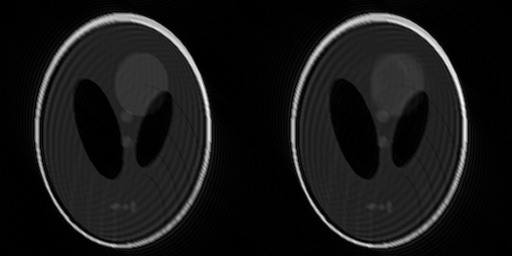}
\plotleftsmall{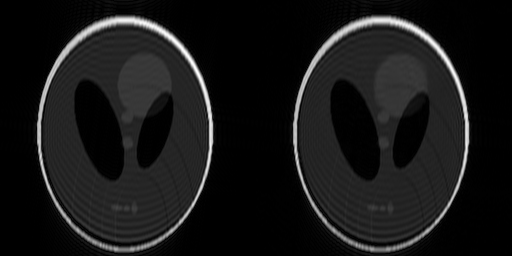}
\caption{\textbf{Dynamic dataset: example training images.} We show a sequence of five consecutive frames from our dynamic training dataset, where the topmost interior ellipse moves over time (from left to right) at the same time as the measurement angles rotate repeatedly through the five PROPELLER blades.\vspace{-0.3cm}} \label{fig:dynamic_input}
\end{figure}

This dataset simulates key aspects of real dynamic PROPELLER MRI, notably that we never observe the entire Fourier domain at any individual timestep. In \Cref{fig:dynamic_input} we show a sequence of five consecutive images in this dynamic dataset. Note that the top ellipse shifts to the right while the measurement blades rotate through $k$-space, producing different chemical shift and undersampling blur at each timestep. During the entire 67-frame sequence, the moving ellipse completes 2.5 periods of its cyclic motion.

\section{Experimental Results} \label{sec:results}

Our quantitative results are summarized in \Cref{tab:results}. The baseline is constructed by taking the five $k$-space PROPELLER blade measurements (or the five most recent blades in the dynamic setting), computing a coverage-weighted average over them in $k$-space, then computing the inverse DFT of the result. Visual comparisons are also provided for each dataset, as well as fat--water separation and a visual inspection of our learned model parameters.

\begin{table}[h]
\begin{tabular}{l|cccc}
         & Static & Static & Static & Dynamic \\ 
         & Synthetic & Liver & Breast & Synthetic \\ \hline
Baseline &       15.74           &      20.38        &      25.98         &     17.08 \\ 
Ours     & \textbf{34.53}            & \textbf{38.66}        & \textbf{41.43}         &   \textbf{37.21}              
\end{tabular}
\caption{\textbf{Quantitative results.} We compare PSNR values for our method and a baseline, across four synthetic and real datasets.\vspace{-0.3cm}}
\label{tab:results}
\end{table}

\paragraph{Static Shepp--Logan dataset.}
Our static synthetic results are shown in \Cref{fig:static_shepp_result}, in which we compare our reconstruction of the classic Shepp--Logan phantom with a simple baseline reconstruction. The baseline recovers the basic structure of the Shepp--Logan phantom, but suffers severe off-resonance artifacts most visible around the bright outer ellipse, which simulates fatty tissue that is most susceptible to chemical shift. Our reconstruction is free of these off-resonance effects and successfully combines the five undersampled blade measurements into a single clean image.

\newcommand{\plotleftthird}[1]{%
  \adjincludegraphics[trim={{0\width} {0\height} {0.5\width} {0\height}}, clip, width=0.32\linewidth]{#1}%
}
\newcommand{\plotrightthird}[1]{%
  \adjincludegraphics[trim={{0.5\width} {0\height} {0\width} {0\height}}, clip, width=0.32\linewidth]{#1}%
}
\newcommand{\plotallthird}[1]{%
  \adjincludegraphics[trim={{0\width} {0\height} {0\width} {0\height}}, clip, width=0.32\linewidth]{#1}%
}
\newcommand{\plotallfull}[1]{%
  \adjincludegraphics[trim={{0\width} {0\height} {0\width} {0\height}}, clip, width=\linewidth]{#1}%
}
\begin{figure}[h]
\centering
\plotleftthird{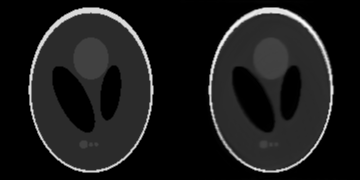}
\plotallthird{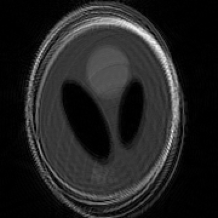}
\plotrightthird{figures/static_good.png}
\plotallfull{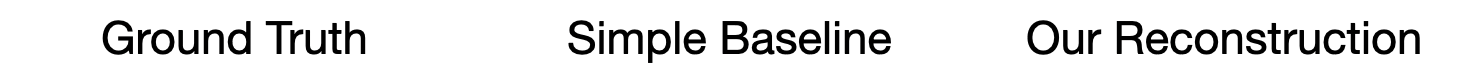}
\caption{\textbf{Static synthetic dataset: results.} We compare our reconstruction with both the ground truth and a simple baseline reconstruction that directly combines the five PROPELLER blade measurements in $k$-space. The baseline is highly susceptible to off-resonance artifacts particularly of the ``fatty'' outer ellipse, whereas our method successfully resolves the reference image.\vspace{-0.3cm}} \label{fig:static_shepp_result}
\end{figure}

\begin{figure}[h]
\centering
\plotallfull{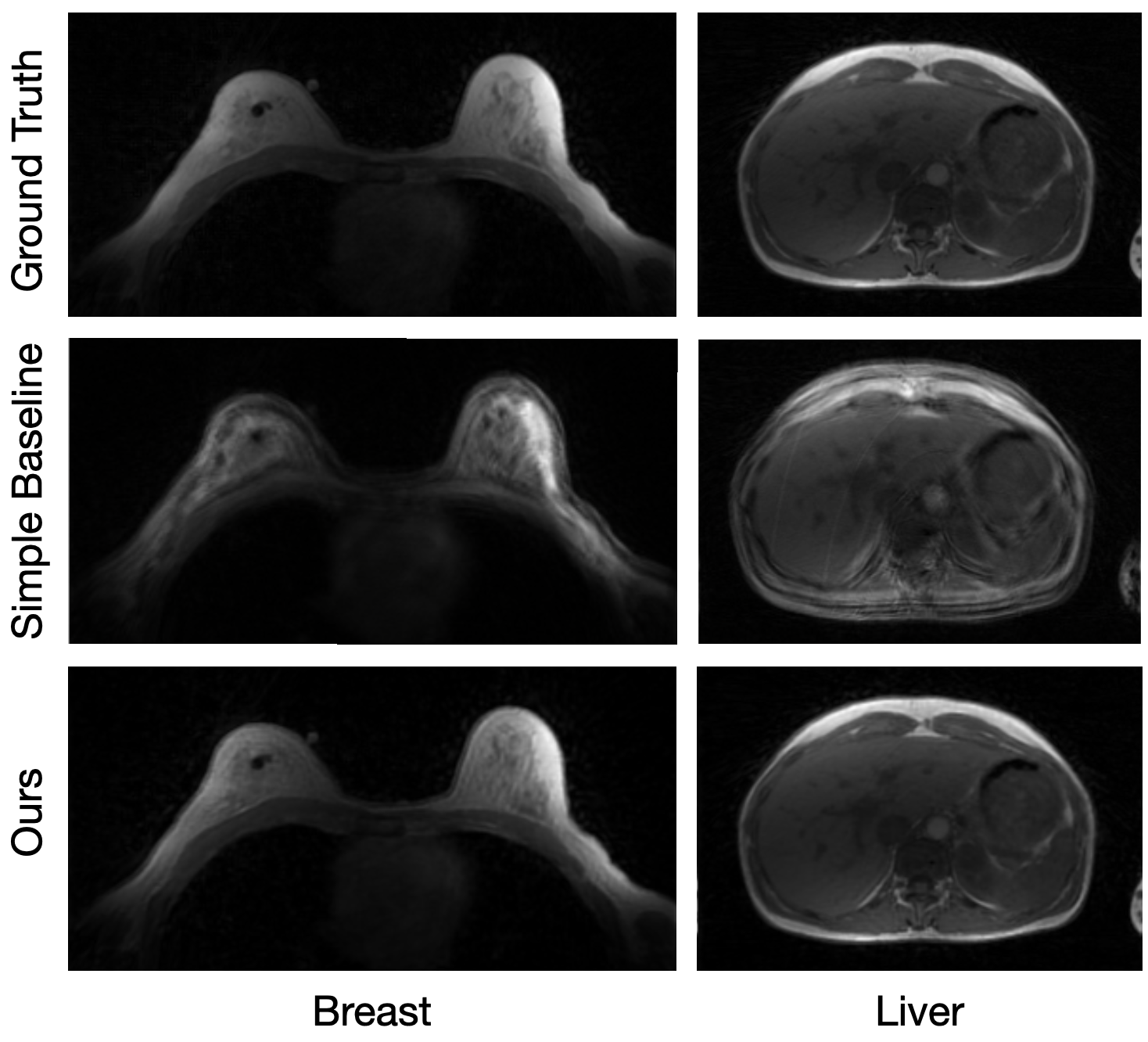}
\caption{\textbf{Static real datasets: results.} For each dataset, we compare our reconstruction with both the ground truth and a simple baseline reconstruction that directly combines the five PROPELLER blade measurements in $k$-space. The baseline is highly susceptible to off-resonance artifacts of fatty tissues, whereas our reconstruction successfully resolves the reference image.\vspace{-0.5cm}} \label{fig:static_real_result}
\end{figure}

\paragraph{Static liver and breast datasets.}
Our real-data static results parallel our results on the Shepp--Logan phantom, and are shown in \Cref{fig:static_real_result}, with the breast cross-section on the left and the liver cross-section on the right. In both cases, we again find that our method cleanly recovers the ground truth image without the off-resonance artifacts present in the baseline reconstruction.

\begin{figure}[h]
\centering
\plotallfull{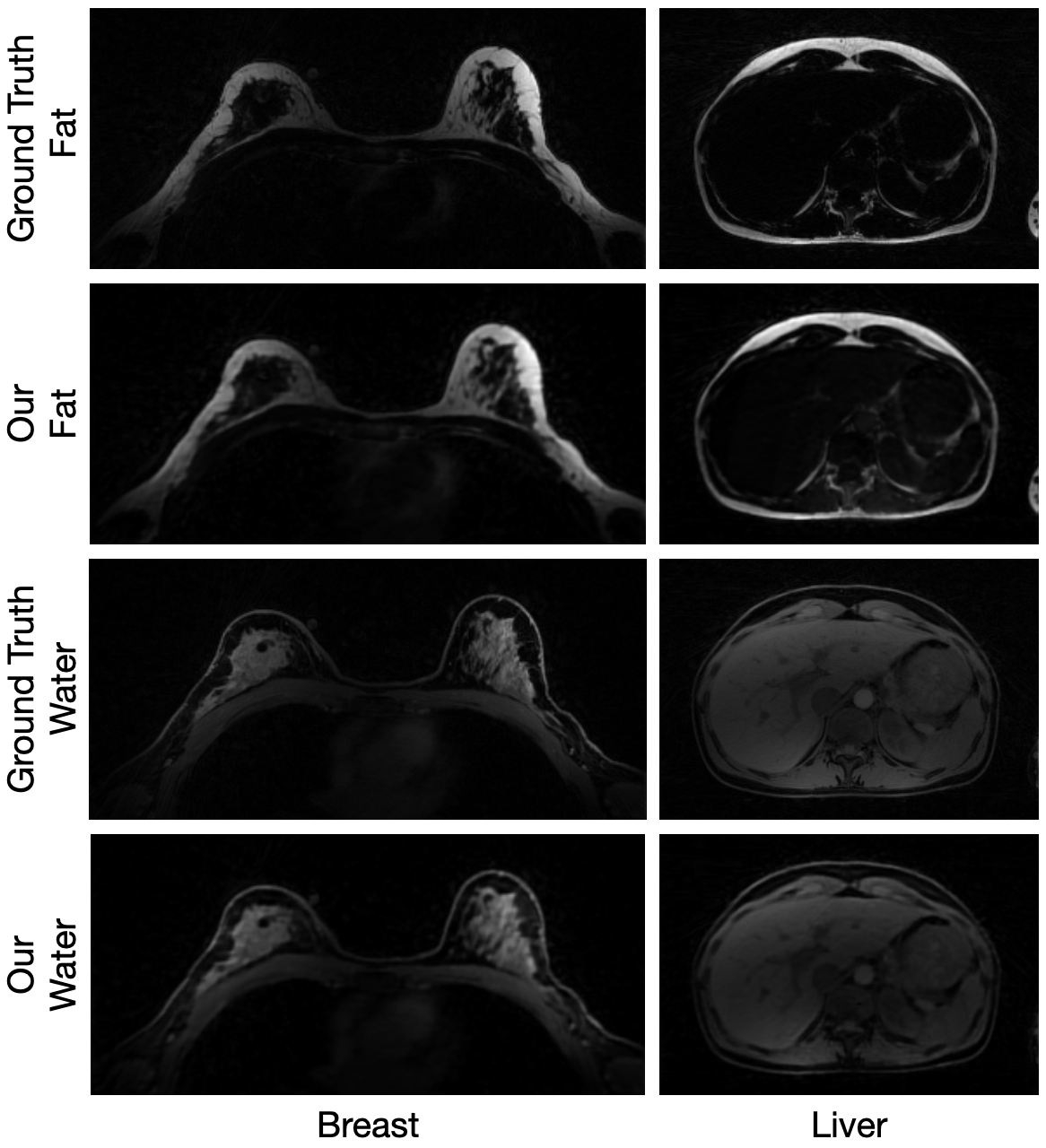}
\caption{\textbf{Fat--water separation: results.} Our method recovers a continuous spectral dimension which can be easily rendered into separate fat and water layers that closely match the reference fat and water images from Dixon-RAVE \cite{dixon-rave}.\vspace{-0.5cm}} \label{fig:fat_water_separation}
\end{figure}

\paragraph{Fat--water separation.}
Because our method recovers a continuous spectral axis, we can render out slices of the final reconstructed image or video that contain only tissues with certain chemical properties. For example, by separating our reconstructed volume into lower and upper portions along the spectral axis, we can render out clinically valuable fat-only and water-only images. We show this chemical separation in \Cref{fig:fat_water_separation} on the real breast and liver datasets, for which we use the Dixon-RAVE \cite{dixon-rave} fat--water separation as ground truth.

\begin{figure}[h]
\centering
\includegraphics[width=\linewidth]{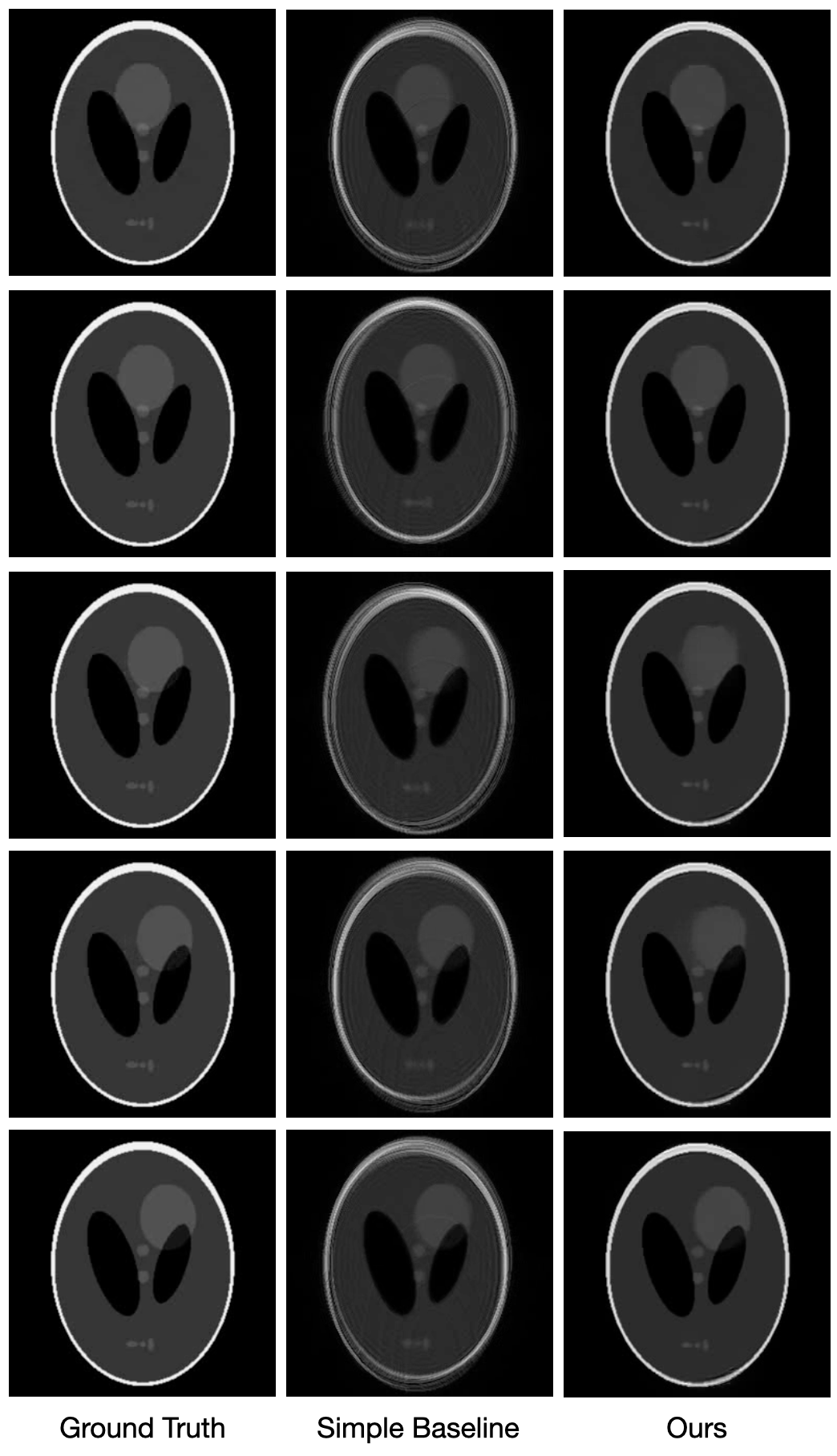}
\caption{\textbf{Dynamic dataset: example reconstructions.} From top to bottom, we show a sequence of five consecutive frames from our dynamic experiment. For each frame, we compare our reconstruction with a simple baseline and the ground truth. 
\vspace{-0.3cm}} \label{fig:dynamic}
\end{figure}

\paragraph{Dynamic Shepp--Logan dataset.}
We summarize the results of our dynamic experiment in \Cref{fig:dynamic}.
Our dynamic baseline method is similar to our static baseline, except that for each frame's reconstruction we use the previous five PROPELLER blades---an exact tiling over $k$-space---in the coverage-weighted average and inverse DFT. This results in a video that lags slightly behind the ground truth as well as suffering from off-resonance (chemical shift). Our dynamic reconstruction recovers the motion of the moving ellipse while correcting for off-resonance and PROPELLER blade undersampling blur.

\begin{figure}[h]
\centering
\includegraphics[width=\linewidth]{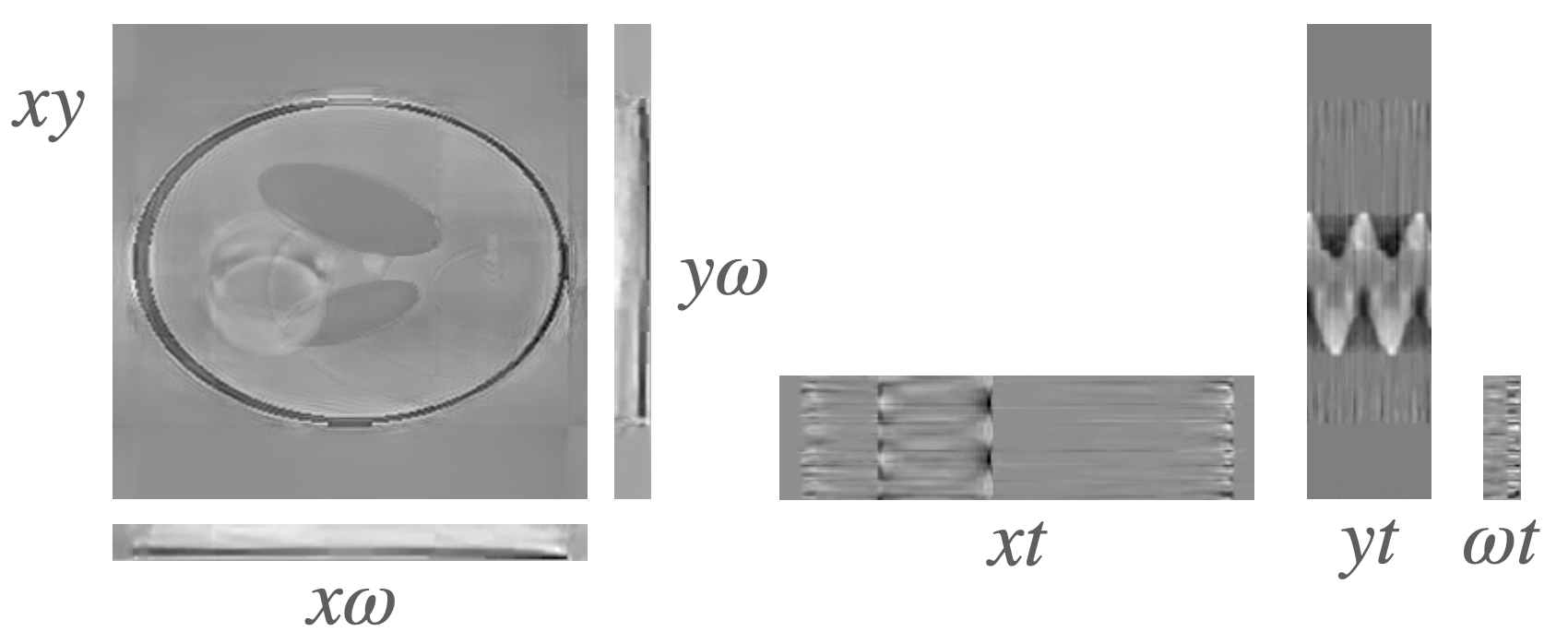}
\caption{\textbf{Dynamic dataset: model parameters.} Our dynamic volume model consists of six grids representing every pairing of the coordinates $x$, $y$, $t$, and $\omega$. By visualizing the learned parameters (averaged along the feature dimension), we can see the basic Shepp--Logan spatial structure in the $xy$ grid, evidence of fat--water separation in the $x\omega$ and $y\omega$ grids, and the recovered motion in the $xt$ and $yt$ grids.\vspace{-0.3cm}} \label{fig:planes}
\end{figure}

\paragraph{Model interpretation.}
Our grid-based K-Planes model lends itself to interpretation, as shown in \Cref{fig:planes}. The learned parameters show recovery of the spatial structure of the Shepp--Logan phantom, the sinusoidal motion of the moving ellipse, and the fat--water separation along the spectral dimension.

\section{Discussion}

In this work, we propose a strategy to resolve off-resonance artifacts in both static and dynamic PROPELLER MRI, leveraging a mathematical equivalence between off-resonance artifacts and the projection effects ubiquitous in computer vision and graphics. Our method works directly from raw gradient-echo PROPELLER blade measurements without any additional field map estimation or pretraining data, and recovers both the spatiotemporal structure as well as spectral properties of the tissue, including fat--water separation. It removes off-resonance or chemical shift artifacts while resolving motion to produce a video reconstruction. We demonstrate promising results on both static and dynamic synthetic datasets, as well as two real static abdominal datasets measuring cross-sections through breast and liver tissue. 

\paragraph{Limitations.}
The primary limitation of our work is that even our experimental liver and breast data was not sampled using gradient-echo PROPELLER MRI, but was instead collected using other $k$-space sampling strategies. We then processed these measurements using our forward model to simulate the measurements we might have collected using a gradient-echo PROPELLER sequence.
Nonetheless, our preliminary results are promising and we hope that followup work will apply and validate our method on real gradient-echo PROPELLER measurements, with the ultimate goal of improving clinical scan time and reconstruction quality. 

Specifically, we note that our forward model uses a Cartesian grid in $k$-space along with uniform DFT and inverse DFT implementations.
In practice, PROPELLER blades are collected along blade-aligned lines rather than axis-aligned lines in $k$-space, and therefore require nonuniform DFT and inverse DFT in the full forward model. This difference is one of convenience and implementation; future work with real data should be able to use our method exactly as described here, except for replacing the uniform DFT and inverse DFT with nonuniform implementations.


\paragraph{Conclusions.}
Our physics-based method enables high-quality magnetic resonance reconstruction using fast and motion-robust gradient-echo PROPELLER measurements, without requiring slower spin-echo measurements, additional field map measurements, or pretraining data. We hope that our method leads to faster scan times and higher temporal resolution of clinically relevant patient motions such as respiration and peristalsis.

\paragraph{Acknowledgments.}
We would like to thank Miki Lustig, Efrat Shimron, Suma Anand, and John Pauly for helpful discussions.
Gordon Wetzstein is supported in part by Stanford HAI and in part by an Army Research Office Presidential Early Career Award for Scientists and Engineers.
Mert Pilanci is supported in part by the National Science Foundation (NSF) under Grant DMS-2134248; in part by the NSF CAREER Award under Grant CCF-2236829; in part by the U.S. Army Research Office Early Career Award under Grant W911NF-21-1-0242; in part by the Stanford Precourt Institute.
Sara Fridovich-Keil is supported by an NSF Mathematical Sciences Postdoctoral Research Fellowship under award number 2303178.
Any opinions, findings, and conclusions or recommendations expressed in this material are those of the authors and do not necessarily reflect the views of the National Science Foundation.

{
    \small
    \bibliographystyle{ieeenat_fullname}
    \bibliography{main}
}


\end{document}